\documentclass[floatfix, ApJL,twocolumn]{aastex631}

\usepackage{amssymb}
\usepackage{amsmath}
\usepackage{aas_macros}
\usepackage{color,units}
\usepackage{xspace}
\usepackage{subfigure}
\usepackage{comment}
\usepackage{xcolor}
\usepackage{bm}
\usepackage{dcolumn}
\usepackage{hyperref}
\usepackage{enumitem} 
\usepackage{array}
\usepackage{nicematrix}
\usepackage{hhline}
\usepackage{graphicx}
\usepackage{multirow,units}
\usepackage[normalem]{ulem}
\usepackage{soul}
\usepackage[math]{cellspace}
\usepackage{nicematrix} 
\newlist{todolist}{itemize}{2}
\setlist[todolist]{label=$\square$}

\newcommand{\acf}{angular correlation function}
\newcommand{\hd}{Hellings-Downs}

\definecolor{ForestGreen}{RGB}{34,139,34}

\newcommand{\SPA}{School of Physics and Astronomy, Monash University, Clayton VIC 3800, Australia}
\newcommand{\OzGravMonash}{OzGrav: The ARC Centre of Excellence for Gravitational Wave Discovery, Clayton VIC 3800, Australia}
\newcommand{\WVU}{West Virginia University Department of Physics and Astronomy, Morgantown, WV, 26501, USA}
\newcommand{\MPI}{Max Planck Institute for Gravitational Physics, Leibniz Universität Hannover,
Callinstrasse 38, D-30167 Hannover, Germany}

\newcommand{\Yale}{Department of Astronomy, 52 Hillhouse Avenue, New Haven, CT 06511}
\newcommand{\BHI}{Black Hole Initiative, 20 Garden Street, Cambridge, MA 02138}

\newcommand{\Swinburne}{Centre for Astrophyics and Supercomputing, Swinburne University of Technology, Hawthorn, VIC, 3122, Australia }
\newcommand{\OzGrav}{OzGrav: The ARC Centre of Excellence for Gravitational Wave Discovery}

\newcommand{\Birmingham}{School of Physics and Astronomy \&	 Institute for Gravitational Wave Astronomy, University of Birmingham, Birmingham, B15 2TT, UK}

\begin{document}
\title{The International Pulsar Timing Array checklist for the detection of nanohertz gravitational waves}

\author{Bruce Allen}\affiliation{\MPI}

\author{Sanjeev Dhurandhar} \affiliation{Inter University Centre for Astronomy \& Astrophysics, Ganeshkhind, Pune - 411 007, India}

\author{Yashwant Gupta}
\affiliation{National Centre for Radio Astrophysics, Pune University Campus, Pune 411007, India}

\author{Maura McLaughlin}\affiliation{\WVU}

\author{Priyamvada Natarajan}
\affiliation{\Yale}
\affiliation{\BHI}

\author{Ryan M. Shannon}
\affiliation{\Swinburne}
\affiliation{\OzGrav}

\author{Eric Thrane}
\affiliation{\SPA}
\affiliation{\OzGravMonash}

\author{Alberto Vecchio}\affiliation{\Birmingham}

\begin{abstract}
\noindent
Pulsar timing arrays (PTAs) provide a way to detect gravitational waves at nanohertz frequencies. In this band, the most likely signals are stochastic, with a power spectrum that rises steeply at lower frequencies.  Indeed, the observation of a common red noise process in pulsar-timing data suggests that the first credible detection of nanohertz-frequency gravitational waves could take place within the next few years.  The detection process is complicated by the nature of the signals and the noise: the first observational claims will be statistical inferences drawn at the threshold of detectability. To demonstrate that gravitational waves are creating some of the  noise in the pulsar-timing data sets, observations must exhibit the Hellings and Downs curve---the angular correlation function associated with gravitational waves---as well as demonstrating that there are no other reasonable explanations.  To ensure that detection claims are credible, the International Pulsar Timing Array (IPTA)  has a formal process to vet results prior to publication. This includes internal sharing of data and processing pipelines between different PTAs, enabling independent cross-checks and validation of results. To oversee and validate any detection claim, the IPTA has also created an eight-member Detection Committee (DC) which includes four independent external members. IPTA members will only publish their results after a formal review process has concluded. 
This document is the initial DC checklist, describing some of the conditions that should be fulfilled by a credible detection.
At the present time none of the PTAs have a detection claim; therefore this document serves as a road map for the future.
\end{abstract}

\section{Introduction}
A nanohertz-frequency stochastic background of gravitational waves creates corresponding low-frequency fluctuations in pulsar timing residuals \citep{Sazhin,Detweiler,Foster-Backer:1990}. 
Such fluctuations may first be inferred with the same spectrum in
all PTA pulsars \cite[][]{Hobbs+09}.  If this common-spectrum process arises from a gravitational-wave background, it will have statistically consistent amplitude and spectral shape in every pulsar.
However, the intrinsic rotation noise of pulsars, and the noise generated by pulse propagation through the interstellar medium (ISM) are not fully understood \cite[][]{Groth75,Shannon+10,Melatos+14}.  This means that the observation of a common (red) noise process is not, by itself, compelling evidence of gravitational waves.

One way to distinguish gravitational waves from other sources of pulsar timing fluctutations was proposed by \citet{hd83}.  They show that low-frequency gravitational waves create a pattern of angular correlations between pulsars in different parts of the sky. These correlations have quadrupolar signature described by the Hellings and Downs curve \citep{hd83}. 
A clear observation of this pattern of angular correlations is one way to distinguish a gravitational-wave background from intrinsic pulsar spin noise, interstellar-medium effects , observatory clock errors, ephemeris errors, and other sources of noise \cite[][]{Tiburzi+16}.

For several decades, PTAs have placed increasingly stringent constraints on the amplitude of the nanohertz-frequency  gravitational-wave background through spectral analyses of pulsar timing measurements. 
Recently, three PTA collaborations---the European Pulsar Timing Array \cite[EPTA,][]{Kramer+13}, the North American Nanohertz Observatory for Gravitational Waves \cite[NANOGrav,][]{McLaughlin13}, and the Parkes Pulsar Timing Array \cite[PPTA,][]{Manchester+13} all reported the detection of a common-spectrum process in their current datasets \citep{nanograv2020,epta2021,ppta2021}. Here, ``common spectrum" means that the fluctuations in different pulsar arrival times are described by the same spectrum, but does \emph{not} imply that the fluctuations have a common source (in which case they would be correlated between different pulsars).
Indeed, this same common-spectrum process is also detected in the most recent International Pulsar Timing Array (IPTA) dataset, consisting of data from these three regional collaborations  \citep{ipta2022}.

The measured amplitude and spectral index of this common-spectrum process is consistent in  all four datasets at the 2$\sigma$ level \citep{ipta2022}
It is also consistent with some theoretical predictions for a gravitational-wave background from a cosmological population of supermassive black hole binaries \cite[e.g.,][]{Begelman+80,Rajagopal+95, Sesana13}, though, given the current level of understanding of the uncertainties in the modeling of the growth and evolution of the supermassive black hole population,a wide range of predictions have been made \cite[e.g.,][]{Sesana+08,Ravi+12,Zhu+19}. 
However, angular correlations with the characteristic quadrupolar signature have not yet been published. Therefore, so far, it has not been possible to conclude that the common-spectrum process is due to an astrophysical/cosmological  gravitational-wave background. 
However, if the currently observed common process is due to gravitational waves, simulations predict that the spatial correlations could be detectable at the 4--5$\sigma$ level in the NANOGrav 15-yr dataset \citep{pol2021} and similar significance detections may be soon possible in other PTA data sets.

In anticipation of evidence for a gravitational-wave background in one or multiple PTA datasets currently under analysis, the IPTA collaboration (which includes the EPTA, NANOGrav, PPTA and the Indian Pulsar Timing Array) convened a Detection Committee comprising of members from all four PTAs, and four independent experts. The Detection Committee is tasked with developing a detection checklist to help verify candidate gravitational-wave signals. This document is the result of the work of that Committee.
At the present time none of the PTAs have a detection claim.
This checklist has not been applied to the submitted papers, which therefore have not been vetted by the Detection Committee.

\section{Detection checklist}
The detection checklist consists of three subsections.
In \ref{significance}, we describe checklist items designed to ensure that Hellings-Downs correlations are statistically significant.
In \ref{checks}, we describe checklist items that provide consistency checks to ensure that the signal and noise model are well specified, and that the signal cannot be easily explained by systematic error.
Finally, in \ref{vetting}, we describe checklist items that ensure that results are cross-checked and vetted by independent experts.

\subsection{Statistical significance}\label{significance}
\begin{todolist}
\item \textbf{The Hellings-Downs signal is evident in the data with $5\sigma$ significance.}
    There are different ways to show this.
    We provide two examples: one Bayesian and one frequentist.  

    \textit{Bayesian formulation.}
    Construct the $5\sigma$, highest posterior density credible interval for the gravitational-wave amplitude $A$.
    Show that this credible interval excludes $A=0$.
    The posterior is constructed using only cross-power 
    (no auto-power) so that this detection statement is not influenced by the presence of (quasi-) common red noise, which may or may not be due to gravitational waves.
    When constructing this posterior, analysts should marginalize over all relevant sources of uncertainty including pulsar noise models and astrophysical uncertainty in the signal model.    

    \textit{Frequentist formulation.}
    Construct an estimator for the amplitude of the Hellings-Downs amplitude $Y_\text{HD}$ with associated uncertainty $\sigma_Y$ using only point estimates for the angular correlation function from pairs of distinct pulsars and the associated error bars.
    The following is a frequentist detection statistic for the Hellings-Downs signal:
    \begin{align}
        \rho_\text{HD} = Y_\text{HD}/\sigma_Y .
    \end{align}
    Calculate the null distribution of $\rho_\text{HD}$ under the assumption that the signal and noise models are correctly specified and that no correlation is present.
    The observed value of $\rho_\text{HD}$ occurs with a probability of $p\leq 3\times10^{-7}$ in the null distribution ($\geq 5\sigma$ confidence).
    
    \item \textbf{The statistical significance computed in the previous bullet is consistent with estimates from bootstrap methods.}
    At least $N=1000$ (preferably more) quasi-independent noise realisations from phase-scrambling and/or sky-scrambling have been analysed.\footnote{These boot-strap noise realisations should take into account the relative quality of each pulsar as per \cite{Cornish2016}.} 
    Evidence needs to be provided to demonstrate that the required number of quasi-independent realisations has been achieved. 
    Repeating the previous test, but with boot-strap noise realizations, yields a detection that is consistent with a $\geq 5 \sigma$ detection. 
    There are no instances of boot-strap noise with false-positive detections that are more significant than the signal present in the data.\footnote{In the Bayesian formulation, there are no boot-strap realizations where $A=0$ is excluded with higher credibility than it is in the data. In the frequentist formulation, there are no boot-strap realizations with larger values of $\rho_\text{HD}$.}
\end{todolist}

\subsection{Consistency checks}\label{checks}
The data must be consistent with the signal and noise models.
The checklist items in this section are designed to ensure that the angular correlation function is consistent with a Hellings-Downs signal and clearly quadrupolar in nature. 
It also needs to be demonstrated that the significance is not overly reliant on a small number of pulsars

\begin{todolist}
    \item \textbf{The \acf{} is consistent with the \hd{} curve.}
    Compute a $p$-value under the null hypothesis that the measured angular correlation function is consistent with the \hd{} curve for an isotropic stochastic background. 
    It can be calculated using a $\chi^2$ statistic, which sums over pulsar pairs (or angular bins):
    \begin{align}\label{eq:chisq}
        \chi^2(h) = & (y_\alpha - h^2\mu_\alpha)^\dagger \, C^{-1}_{\alpha\beta}(h) \, (y_\beta - h^2\mu_\beta) .
    \end{align}
    Repeated indices imply summation.
    Here, $y_\alpha$ is the cross correlation for pulsar pair $\alpha$, $\mu_\alpha$ is the Hellings-Downs curve evaluated at the angular separation for pair $\alpha$, $h$ is the gravitational-wave background amplitude, and $C_{\alpha\beta}$ is the covariance matrix.
    The value of $h^2$ may be set ``externally," i.e., based on autocorrelation measurements, or ``internallly" by minimizing $\chi^2$, which reduces the number of degrees of freedom by one.
    The  $\chi^2$ value is used to calculate a $p$-value.
    If the data are consistent with the \hd{} hypothesis, we expect $p\approx 50\%$.
    A small $p$-value $<0.02$ indicates that the model does not provide an adequate fit.
    
    In principle, the covariance matrix in Eq.~\ref{eq:chisq} incorporates the effects of measurement uncertainty arising from pulsar and measurement noise and cosmic variance arising from the random amplitudes, phases and locations of gravitational-wave sources; see \cite{AllenRomano}.
    If it is included, the cosmic variance contributions should be consistent with studies published in peer-reviewed journals.
    
    It is useful to contrast this checklist item (related to goodness of fit) with the checklist items in \ref{significance} related to significance.
    We provide examples in Table~1 showing different scenarios.
    The first rows illustrate two different failure modes while the final row provides an example of a detection claim that passes.

\begin{widetext}
\begin{table*}
\begin{tabular}{ |p{3.5cm}||p{3cm}|p{3cm}|p{3cm}| }
    \hline
 \textbf{Example situation} & \textbf{Significance}: Credibility with which we exclude $A=0$ & \textbf{Consistency}: $p$-value under the null hypothesis (that the data are described by the model) & \textbf{pass/fail} \\\hline
Significant, but misspecified & {\color{ForestGreen}$1.32\times10^{-7}$} & {\color{red}$0.003$} & {\color{red}\textbf{fail}} \\\hline
Not sufficiently significant, but adequately specified & {\color{red}$3.25\times10^{-3}$} & {\color{ForestGreen}0.67} & {\color{red}\textbf{fail}} \\\hline
Significant, adequately specified   & {\color{ForestGreen}$1.32\times10^{-7}$} & {\color{ForestGreen}$0.67$} & {\color{ForestGreen}\textbf{pass}} \\\hline
\end{tabular}
\caption{Example situations illustrating how a detection claim is expected to pass or fail based on the statistical significance of the signal and the consistency of the data with the model. A failure is marked with red while a pass is marked with green.
}
\end{table*}
\label{tab:examples}
\end{widetext}

    \item \textbf{The signal is clearly quadrupolar.}
    In particular, the pure \hd{} correlation is preferred over a model consisting of a pure monopolar correlation and/or a pure dipolar correlation with a Bayes factor of $>100$.
    If ``quadrupole + monopole'' or ``quadrupole + dipole'' hypotheses are significantly preferred over the pure \hd{} hypothesis (with Bayes factor $\gtrsim 100$), then there are likely still artifacts in the data that are not yet correctly modeled (e.g., clock errors, ephemeris errors, etc.). 
    In such cases, more work may be required before we can establish a detection.
    
    \item \textbf{The signal is present in more than a small number of pulsars.}
    In particular, the number of effective pulsar pairs $n_\text{eff}$ is at least $20$.
    The effective number of pulsar pairs is
    \begin{align}
        n_\text{eff} = &
        \frac{
        \left(\sum_{k=1}^m w_k\right)^2
        }{
        \sum_{k=1}^m w_k^2
        } \nonumber\\
        = & \frac{
        \left(\sum_{k=1}^m \widehat\sigma_k^{-2}\right)^2
        }{
        \sum_{k=1}^m \widehat\sigma_k^{-4}
        } .
    \end{align}
    Here, $w_k$ is the ``weight'' of pulsar pair $k$ while $\widehat\sigma_i$ is the \acf{} uncertainty for pulsar pair $k$.
    The total number of pulsar pairs is $m$.
    (We use the hat to differentiate the uncertainty associated with a pulsar pair $\widehat\sigma_k$ from the uncertainty associated with a cosine angular separation bin $\sigma_i$.)
    Since some pairs are more important than others, $n_\text{eff} < m$.
    If $n_\text{eff} < 20$, then the measurement is dominated by a small number of pulsars such that the \hd{} curve will not be clearly evident through visual inspection of the \acf.
    This check ensures that the \hd{} correlation is visible when the data are binned.
    At least seven well-timed pulsars are required to fulfill this requirement.
    This item is similar in spirit to the ``drop-out factors,'' which pulsar timing arrays use to quantify the relative importance of different pulsars.

    \item \textbf{The result is consistent with previously published analyses by the same PTA.}
    The inferred gravitational-wave background is consistent with previous upper-limit papers written using a subset of the currently available data. If the inferred gravitational-wave background is inconsistent with previously published papers, there is a convincing explanation to account for the discrepancy.
\end{todolist}

\subsection{Independent vetting of results}\label{vetting}
In addition to the checklist items described above, it is useful for the analysis to be independently  verified by domain experts. The following checklist items are designed to facilitate this.

\begin{todolist}
    \item \textbf{The detection team will make available to the other PTAs everything required to reproduce the detection.}
    This includes:
    \begin{itemize}
    \item Standard \texttt{tempo2}-format data including TOA \texttt{tim}  files and ephemerides  \texttt{par} files.
    \item The noise model in either \texttt{Enterprise} or \texttt{temponest} format.
    \item A technical note detailing all the assumptions needed to reproduce the detection.
    \item Weekly ``office hours'' to field questions from other analysis teams.
    \end{itemize}
    The Detection Committee may request additional code or documentation in order to support the reproduction and verification of their main results.

    The process by which NANOGrav, EPTA, and PPTA coordinate their papers is referred to as ``the 3P+ framework.''
    As per 3P+ rules, shared data may not be used for any published papers without permission; it is \emph{only} for checking. Analysis teams will provide data for these checks as soon as practical as part of the 3P+ process. We recommend that the IPTA data combination working group coordinate the sharing of data products and expertise between the constituent PTAs.
    We note that the data combination working group has started some of this work in preparation for Data Release 3. 
    
    Members of the IPTA will have \textit{at least} six weeks from a detection/evidence claim by any PTA to inspect the data. The data products include TOAs, pulsar ephemerides, preferred noise models, and, if possible, intermediate data products such as posterior chains. 
    
    During this time, independent teams are asked to interact with the Detection Committee and to submit brief reports (and supporting analysis materials, as far as practical) to the Detection Committee stating whether or not they are able to find an alternative explanation that could explain the data at least as well as the gravitational-wave hypothesis.
    The independent teams are encouraged to share preliminary findings early with the Detection Committee.
    These reports (and supporting materials) will be made available to the IPTA via the IPTA 3P+ Committee.
    The Detection Committee may request the detection team to provide a reply to one or more reports.
    The Detection Committee will share all reports it receives with the 3P+ Committee.
    The Detection Committee may amend the checklist based on developments during this period.
\end{todolist}

\subsection{Initiating the detection procedure}
The procedure for a PTA to initiate a detection/evidence claim is as follows:
\begin{enumerate}
    \item \textbf{Prepare a paper draft} and send it to the Detection Committee.
    \item \textbf{Prepare a ``response to the detection checklist'' document} and send it to the Detection Committee. This technical note should address every item in the detection checklist provided above. The reply should be relatively succinct for readability; less than one page per checklist item. If multiple pages of supporting material are required for a single checklist item, some of it can be placed in an appendix.
    \item \textbf{Present the detection case to the Detection Committee.}
    The presentation will be scheduled approximately two weeks after the Detection Committee has receives both the paper draft and the response to the detection checklist. This will provide the Detection Committee sufficient time to read both documents and formulate their questions.The Detection Committee may request a follow-up meeting for additional questions and/or to discuss additional investigations that may be deemed necessary.
    \item \textbf{The Detection Committee issues a recommendation} stating whether or not they endorse the detection claim.
    The Detection Committee will aim to provide this recommendation within one month of the final presentation and/or completion of any additional investigations.
    The recommendation will be written as a technical note. If the detection claim is not endorsed, the Detection Committee will specify the aspect that was unconvincing.
    
    The Detection Committee shall endeavour to reach a consensus opinion. However, if that proves impossible, the Committee will draft a report based on the majority opinion. The perspective of the dissenting minority will be included as an appendix.
    In the event of a tie vote, the majority opinion shall be determined by the Detection Committee Chair.
    This report shall be made publicly available.
\end{enumerate}

\appendix
\section{Future work}

The checklist above reflects the short-time scale over which the first paper(s) from the current analyses may become available and the wish of PTA teams to submit them quickly for publication. Taking these factors into account the IPTA Detection Committee has distilled down a list of absolutely necessary checks, which are presented above, but has discussed a much longer list of additional checks that it would strongly encourage PTAs to carry out (partially or in full). Many suggestions have also been received  by the Detection Committee upon circulation of the detection checklist for comments from the IPTA. A revised and updated checklist may well incorporate more detailed checks (e.g.,  marginalizing over different solar system ephemerides, for instance).

\bibliography{refs}

\end{document}